\documentclass{PoS}

\title{NN potentials from IR chiral EFT}

\ShortTitle{NN potentials from IR chiral EFT}

\author{\speaker{Renato Higa}%
         \thanks{This work was partially supported by the BMBF under 
contract number 06BN411.}\\
        Universitity of Bonn\\
        E-mail: \email{higa@itkp.uni-bonn.de}}

\abstract{
Chiral perturbation theory is nowadays a well-established approach 
to incorporate the chiral constraints from QCD. 
Nevertheless, for systems involving one baryon, the power counting 
which dictates the chiral order of observables is not as simple and 
consensual as in the purely mesonic case. 
The heavy baryon approach, which relies on a non-relativistic expansion 
around the limit of infinitely heavy baryon, recovers the usual power 
counting but destroys some analytic properties of the scattering amplitude. 
Some years ago, Becher and Leutwyler proposed a Lorentz-invariant 
formulation of chiral perturbation theory that maintains the required 
analytic properties, but at the expense of a less intuitive power counting. 

Aware of the shortcomings of the heavy baryon formalism, the S\~ao Paulo 
group derived the two-pion exchange component of the nucleon-nucleon 
potential in line with the works of Becher and Leutwyler. 
A striking result was that the long distance properties of the potential 
is determined by the specific low energy region of the pion-nucleon 
scattering amplitude where the heavy baryon expansion fails. 
In this talk I will discuss the origin of such failure and how it 
reflects in the asymptotics of the nucleon-nucleon interaction. 
Some results for phase shifts and deuteron properties will be shown, 
followed by a comparison with the heavy baryon predictions. 
}

\FullConference{6th International Workshop on Chiral Dynamics\\
                 July 6-10 2009\\
                 Bern, Switzerland}

\begin{document}

\section{Analyticity constraints} \label{sec:hbprobl}

\begin{figure}[htb]
\begin{center}
\begin{tabular}{lr}
\raisebox{40pt}{
\includegraphics[width=4.5cm]{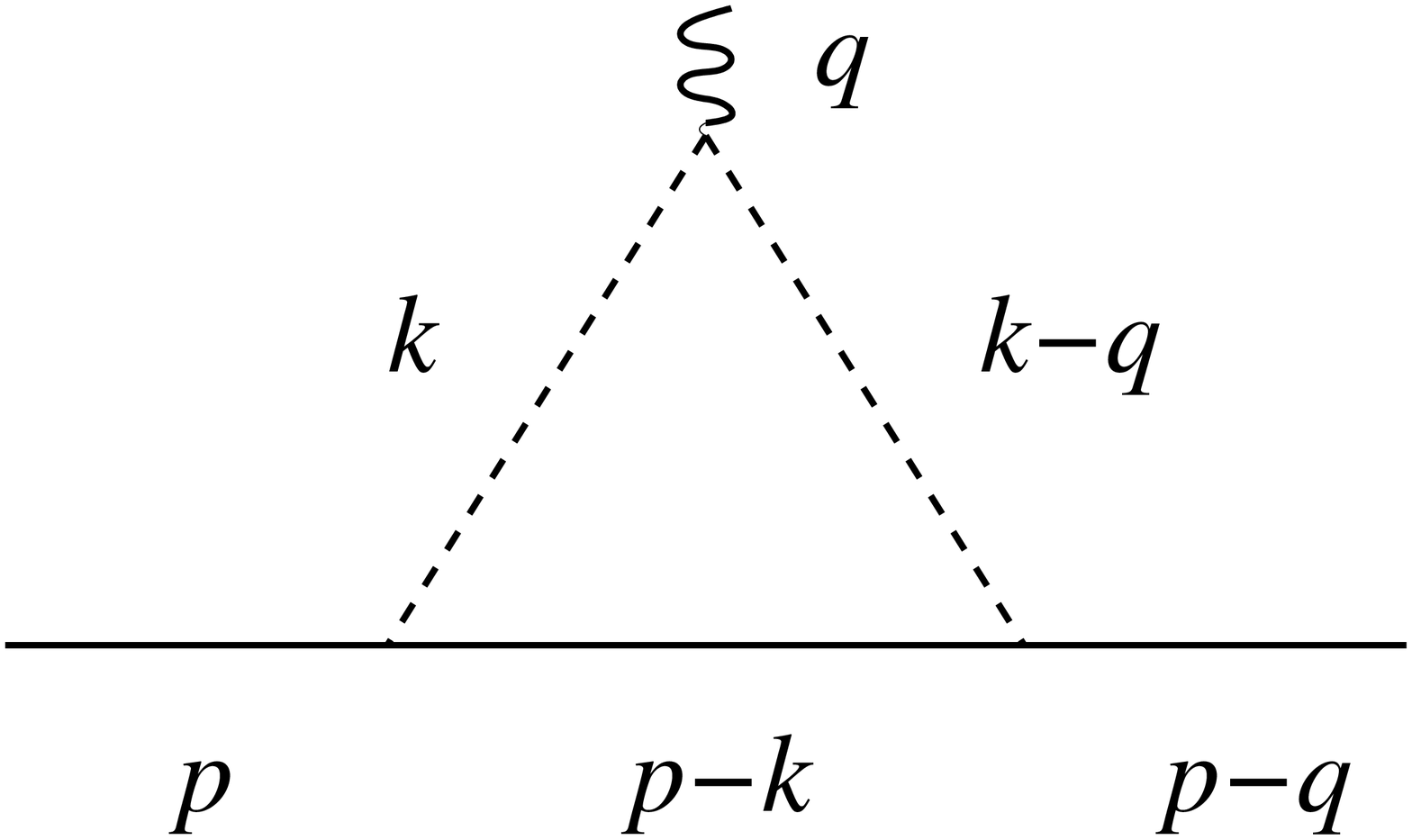} }\hspace{5mm} &
\includegraphics[width=6.5cm,height=5.0cm]{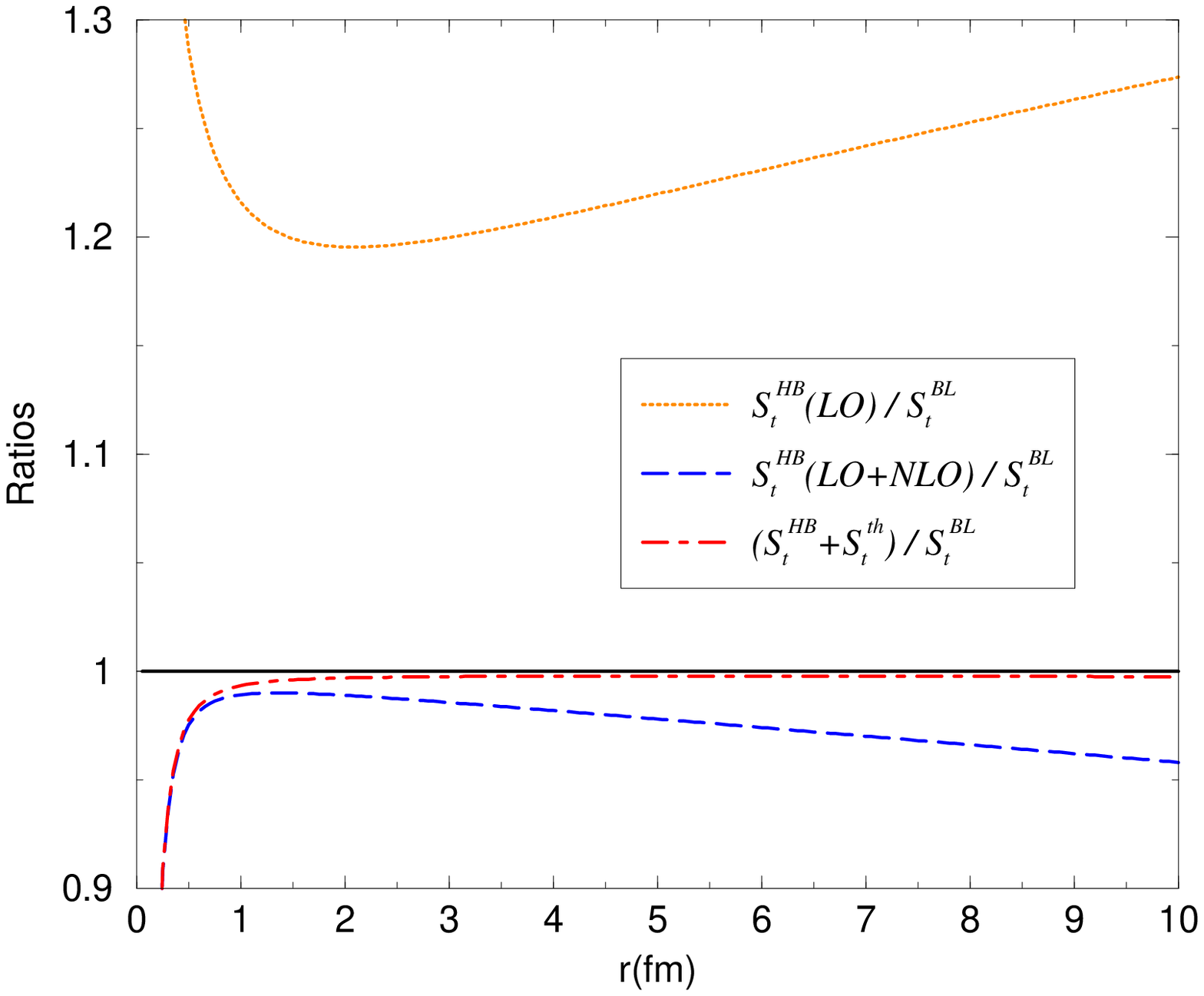}
\end{tabular}
\end{center}
\caption{The triangle diagram that contributes to the nucleon scalar
form factor. On the right panel are shown ratios of the two first 
terms of $\gamma(-q^2)|_{\rm HB}$, divided by $\gamma(-q^2)|_{\rm BL}$. 
For more details, see Ref. \cite{HR03}.}
\label{fig1}
\end{figure}

Becher and Leutwyler \cite{BL} demonstrated that, in the baryon sector, 
convergence of the chiral expansion of certain quantities (form factors, 
for instance) is a delicate issue. This is closely related to the 
presence of an anomalous threshold in the triangle diagram 
(l.h.s. of Fig.\ref{fig1}) for momentum transfer $t$ close to the 
two-pion threshold, $4m_{\pi}^2$. 
Such analytic structure is completely missed in the usual heavy baryon 
(HB) formulation of chiral perturbation theory ($\chi$PT). 
It can only be recovered via a resummation of the HB series to all orders. 
To understand this issue (for more details, see for instance 
Ref. \cite{scherer09}) one starts with the spectral 
representation of the triangle integral,
\begin{equation}
\gamma(t)=\frac{1}{\pi}\int_{4m_{\pi}^2}^{\infty}\frac{dt'}{(t'\!-\!t)}\,
\mbox{Im}\gamma(t')\,,
\quad
\mbox{Im}\gamma(t')\simeq \mbox{Im}\gamma(t')|_{\rm BL}
=\frac{\theta(t'\!-\!4m_{\pi}^2)}{16\pi m_N\sqrt{t'}}
\,\arctan\frac{2m_N\sqrt{t'\!-\!4m_{\pi}^2}}{t'\!-\!2m_{\pi}^2}\,.
\label{eq:triangBL}
\end{equation}
The argument $x=2m_{N}\sqrt{t'-4m_{\pi}^2}/(t'-2m_{\pi}^2)$
is formally counted as order $q^{-1}$ in the HB expansion, which yields 
$\tan^{-1}x= \pi/2 - 1/x + 1/3x^3+\cdots$. In fact, the first two terms 
reproduce the HB result for the triangle graph,
\begin{eqnarray}
\gamma(-q^2)|_{\rm HB}&=&\frac{1}{16\pi^2m_{N}m_{\pi}}
\int_{4m_{\pi}^2}^{\infty}\frac{dt'}{(t'+q^2)}\,\frac{1}{\sqrt{t'}}
\left[\frac{\pi}{2}-\frac{(t'-2m_{\pi}^2)}{2 m_{N}
\sqrt{t'\!-\!4m_{\pi}^2}}\right]
\nonumber\\[2mm]
&=&\frac{1}{16\pi^2m_{N}m_{\pi}}\left[2\pi m_{\pi}\,A(q)+
\frac{m_{\pi}}{m_{N}}\,
\frac{(2m_{\pi}^2+q^2)}{(4m_{\pi}^2+q^2)}\,L(q)\right]\,,
\label{eq:triangHB}
\end{eqnarray}
where $q=|{\mbox{\boldmath $q$}}|$, and $L(q)$ and $A(q)$ are
the usual HB loop functions,
\begin{equation}
L(q)=\frac{\sqrt{4m_{\pi}^2+q^2}}{q}\,\ln
\frac{\sqrt{4m_{\pi}^2+q^2}+q}{2m_{\pi}}\,,
\qquad\qquad
A(q)=\frac{1}{2q}\,\arctan\frac{q}{2m_{\pi}}\,.
\end{equation}
However, it does not take into consideration the case $|x|<1$, where
$t'$ gets closer to $4m_{\pi}^2$. This region controls the long distance
behavior of the triangle diagram, as can be seen by its representation
in configuration space~\cite{HR03,HRR04,RH04},
\begin{equation}
\Gamma(r)= \frac{1}{\pi}\int_{4m_{\pi}^2}^{\infty}dt'\int
\frac{d^3q}{(2\pi)^3}\;e^{-i{\mbox{\boldmath $q$}}\cdot
{\mbox{\boldmath $r$}}}\;\frac{\mbox{Im}\gamma(t')}{t'+q^2}=
\frac{1}{4\pi^2}\int_{4m_{\pi}^2}^{\infty}dt'\;\frac{e^{-r\sqrt{t'}}}{r}\;
\,\mbox{Im}\gamma(t')\,.
\end{equation}
Clearly one sees that, in order to have a good asymptotic 
description of $\Gamma(r)$, one needs a decent representation for 
$\mbox{Im}\gamma(t')$ near $t'=4m_{\pi}^2$, which cannot be provided 
by the usual HB formulation. 

We want to stress that the expansion in $1/m_N$ of our two-pion exchange
nucleon-nucleon potential (TPEP) should, in principle, recover the
expressions from HB$\chi$PT. This can be used as a cross-check of our 
calculation (see next section). However one must keep in mind that,
due to the problem described above, such an expansion is ill-defined 
at long distances and should be avoided.

\section{The Lorentz-invariant two-pion exchange amplitude}

\begin{figure}[!ht]
\begin{center}
\includegraphics[width=5.0in]{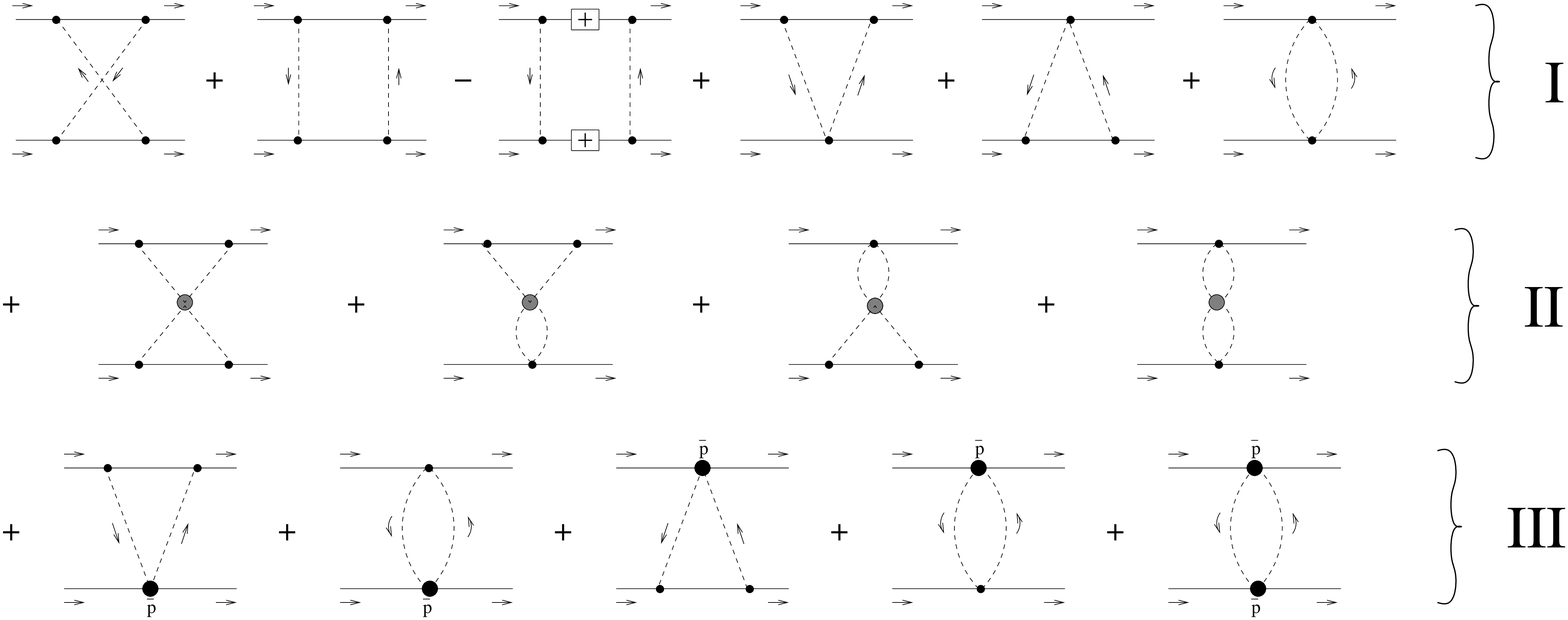}
\end{center}
\caption{Dynamics of the Lorentz-invariant TPEP.}
\label{fig2}
\end{figure}

The two-pion exchange (TPE) amplitude that contributes to the NN 
interaction is closely linked to the $\pi$N amplitude, which is 
generically expressed in terms of two invariant amplitudes 
$D^{\pm}(\nu,t)$ and $B^{\pm}(\nu,t)$,
\begin{eqnarray}
T_{\pi{\rm N}}^{ab}&=&\delta_{ab}\,T_{\pi{\rm N}}^+ 
+ i\epsilon_{bac}\,\tau_c\,T_{\pi{\rm N}}^-\,,
\nonumber\\[2mm]
T_{\pi{\rm N}}^{\pm}&=&\bar u({\mbox{\boldmath $p'$}})
\,\Big[D^{\pm}-\frac{i}{2m_{N}}\,\sigma_{\mu\nu}(p'-p)^{\mu}\,
\frac{(k+k')^{\nu}}{2}\,B^{\pm}\Big]\,u({\mbox{\boldmath $p'$}})\,,
\end{eqnarray}
where ${\mbox{\boldmath $p$}}$ and ${\mbox{\boldmath $p'$}}$
are the initial and final momenta of the nucleon, respectively, while
$\nu=[(p+k)^2-(p-k')^2]/4m_{N}$ and $t=(k'-k)^2$ are the usual Mandelstam
variables. This allows the TPE amplitude to be written as 
\begin{eqnarray}
&&{\cal T}_{\rm TPE}=-\frac{i}{2!}\int[\cdots]
\left[3\,T_{\pi{\rm N}}^{(1)+}\,T_{\pi{\rm N}}^{(2)+}
+2\,{\mbox{\boldmath $\tau$}}^{(1)}\cdot{\mbox{\boldmath $\tau$}}^{(2)}\,
T_{\pi{\rm N}}^{(1)-}\,T_{\pi{\rm N}}^{(2)-}\right]
\\[2mm]
&&=
\left[\bar uu\right]^{(1)}\left[\bar uu\right]^{(2)}{\cal I}_{DD}^{\pm}
-\left[\bar uu\right]^{(1)}
\left[\bar u\,i\,
\sigma_{\mu\lambda}\,\frac{(p^{\prime}\!-\!p)^{\mu}}{2m_{N}}\,u\right]^{(2)}
{\cal I}_{DB}^{\lambda\pm}
-\left[\bar u\,i\,
\sigma_{\mu\lambda}\,\frac{(p^{\prime}\!-\!p)^{\mu}}{2m_{N}}\,u\right]^{(1)}
\left[\bar uu\right]^{(2)}{\cal I}_{DB}^{\lambda\pm}
\nonumber\\[0mm]&&
+\left[\bar u\,i\,\sigma_{\mu\lambda}\,
\frac{(p^{\prime}\!-\!p)^{\mu}}{2m_{N}}\,u\right]^{(1)}
\left[\bar u\,i\,\sigma_{\nu\rho}
\,\frac{(p^{\prime}\!-\!p)^{\nu}}{2m_{N}}\,u\right]^{(2)}
{\cal I}_{BB}^{\lambda\rho\pm}\,,
\label{eqI-TPEstruct}
\end{eqnarray}
where the superscript $(i)$ refers to nucleon $i$. 
The symbol $\int[\cdots]$ represents the four-dimensional
integration with two pion propagators,
\begin{equation}
\int[\cdots]=\int\frac{d^4Q}{(2\pi)^4}\,\frac{1}{(k^2-m_{\pi}^2)}
\,\frac{1}{(k'^2-m_{\pi}^2)}\,,
\end{equation}
with $Q=(k+k')/2$ the average momentum of the exchanged pions. 
The profile functions ${\cal I}$'s are covariant loop integrals written 
in terms of the amplitudes $D^{\pm}$ and $B^{\pm}$,
\begin{equation}
\begin{array}{rclcrcl}
{\cal I}_{DD}^{\pm}&=&-\frac{i}{2}\,\int[\cdots]\,D^{(1)\pm}\,D^{(2)\pm}\,,
&{\mbox{\ \ \ \ }}&
{\cal I}_{DB}^{\lambda\pm}&=&-\frac{i}{2}\,\int[\cdots]\,Q^{\lambda}\;
D^{(1)\pm}\,B^{(2)\pm}\,,\\[2mm]
{\cal I}_{BD}^{\lambda\pm}&=&-\frac{i}{2}\,\int[\cdots]\,Q^{\lambda}\;
B^{(1)\pm}\,D^{(2)\pm}\,,&&
{\cal I}_{BB}^{\lambda\rho\pm}&=&-\frac{i}{2}\,\int[\cdots]
\,Q^{\lambda}Q^{\rho}\;B^{(1)\pm}\,B^{(2)\pm}\,.
\end{array}\label{eqI-prf1}
\end{equation}
With the $\pi$N amplitude to $O(q^3)$ as input \cite{BL}, one generates 
the TPE amplitude to $O(q^4)$, represented graphically by Fig.\ref{fig2}. 
The contributions are grouped in three families of diagrams, according 
to their topology. The first line of Fig.\ref{fig2} corresponds to the 
irreducible one loop graphs with vertices from the $O(q^1)$ $\pi$N 
chiral Lagrangian, ${\cal L}^{(1)}_{\pi{\rm N}}$, with coupling constants 
at their physical values (family I). The second line (family II) 
contains two-loop diagrams with an intermediate $\pi \pi$ scattering, 
while the third line (family III) comprises one loop graphs with 
vertices from ${\cal L}^{(2)}_{\pi{\rm N}}$ and ${\cal L}^{(3)}_{\pi{\rm N}}$, 
as well as one-loop vertex corrections, parametrized in terms of $\pi$N 
subthreshold coefficients. 

The diagram with a plannar box topology (second of family I) 
contains a reducible piece (iterated one pion), which has to be subtracted 
in the definition of the potential. This subtraction is represented by the 
third graph of family I, the symbol ``+'' standing for a nucleon with only its 
positive-energy projection. It is well-known that such projection is not 
uniquely defined \cite{friar99}. In order to compare our results with 
the HB ones we adopt the same subtraction followed by Ref.~\cite{KBW}. 

\section{Comparison with the HB approach}

The expressions for the Lorentz-invariant TPEP at N${}^{3}$LO is lengthy 
and will not be reproduced here (see \cite{HR03,RH04} instead). They are 
written in terms of Lorentz-invariant loop integrals which, due to the 
reasons mentioned in Sec.\ref{sec:hbprobl}, do not admit the naive HB 
expansion. Nevertheless, if one formally performs such unjustified 
expansion, one recovers all but three terms of the HB expressions 
\cite{RH04}. These discrepancies come from two-loop diagrams of family II 
and affect mostly the central isovector component. Numerically, the 
difference is about 10\% up to 3fm and decreases to 5--2\% beyond that. 
It is not significant to the total central isovector component of the TPEP, 
since the contribution of family II is already fairly small \cite{HRR04}.
The origin of such discrepancy still remains to be understood, and it 
might become relevant when more precision is asked for. 

\begin{figure}[!ht]
\begin{center}
\begin{tabular}{ll}
\includegraphics[width=5.5cm]{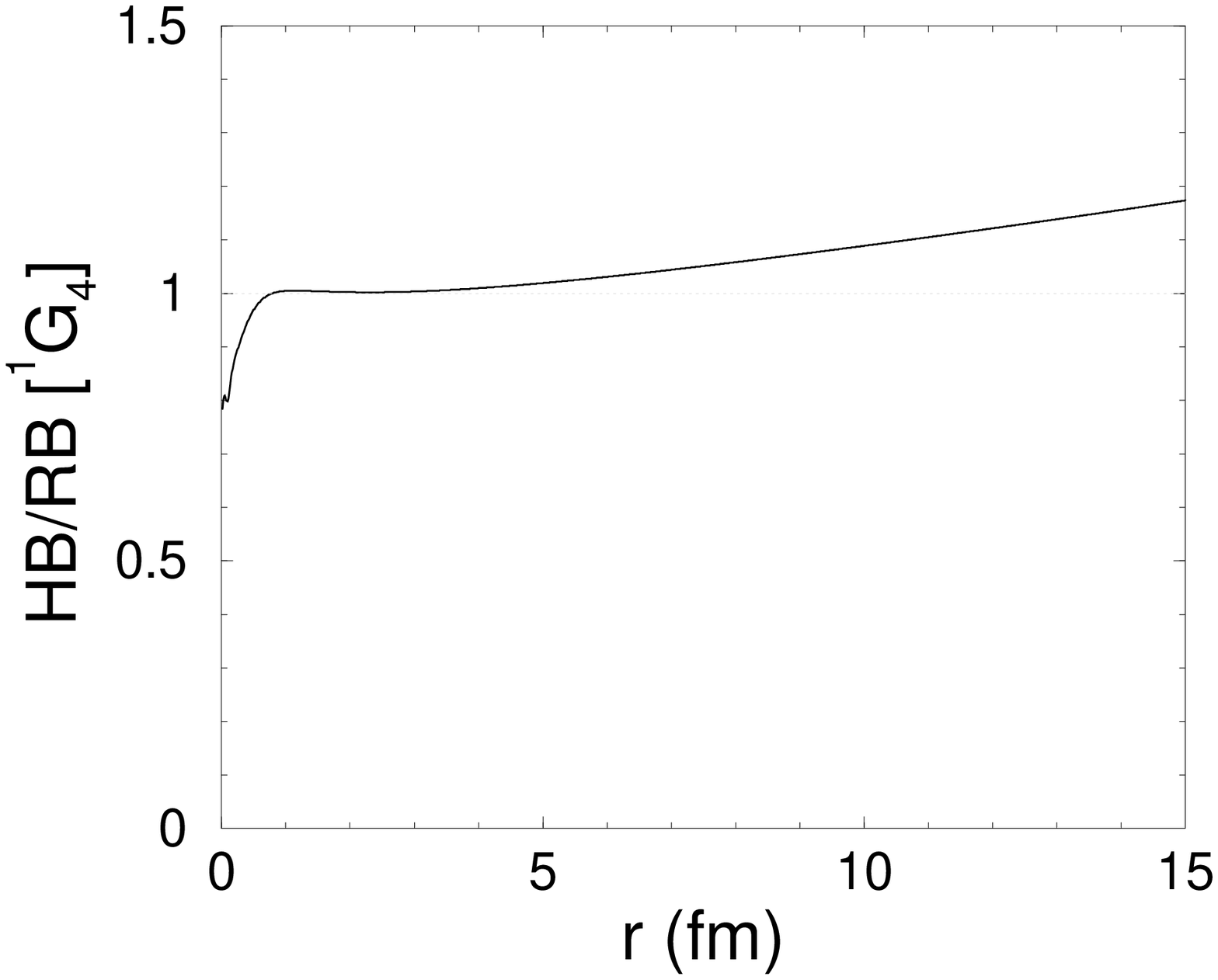}\hspace{5mm}&
\includegraphics[width=5.5cm]{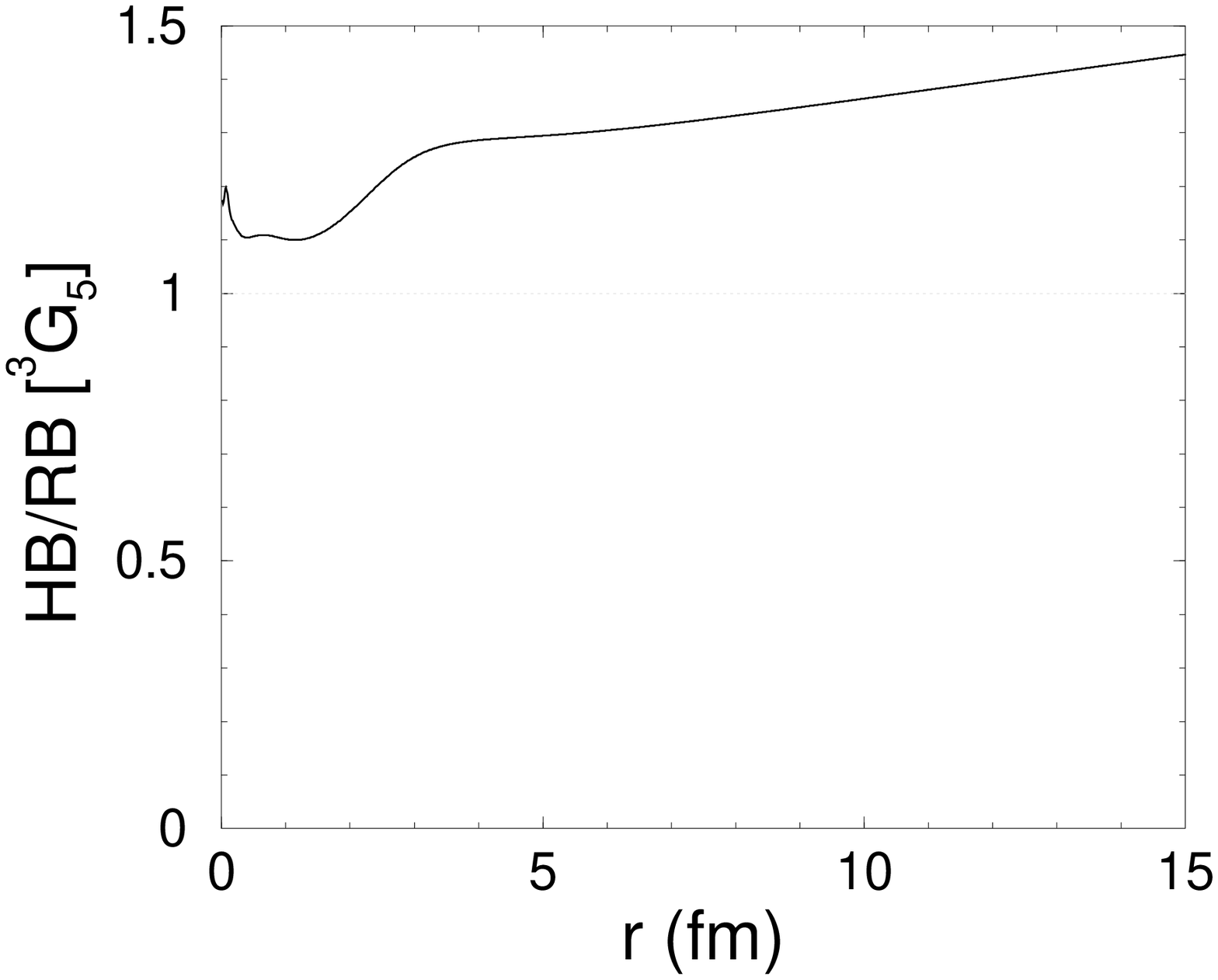}
\end{tabular}
\end{center}
\caption{Ratios between HB and RB formulations of the TPEP, for 
${}^{1}G_4$ and ${}^{3}G_5$ partial waves.}
\label{fig3}
\end{figure}

For now on we ignore the above mentioned discrepancies and concentrate 
only on the effect of expanding (HB) or not (RB) the Lorentz-invariant 
loop integrals. 
In Fig.\ref{fig3} we show the ratio of the HB over the RB versions of 
the TPEP, projected into two specific partial waves. The ${}^{1}G_4$ 
illustrates the general behavior of partial waves with total isospin 
$T=1$, which is similar to the LO+NLO terms in the HB expansion of the 
triangle integral (Fig.\ref{fig1}). On the other hand, the partial waves 
with $T=0$, represented by ${}^{3}G_5$, are more sensitive to the HB 
expansion, with a discrepancy of 10--30\% between 2 and 5fm, and 
increasing to almost 50\% at $r=15$fm. This is due to the 
(isoscalar)$-$3(isovector) structure in those waves, each term having 
nearly the same magnitude. 
Although in Fig.\ref{fig3} we used the set EM 02 from Table \ref{tab1}, 
these qualitative results do not change with the choice of LECs. 

\begin{table}[htb]
\begin{center}
\begin{tabular} {|c|r|r|r|r|r|}
\hline
LEC & BM 00 & Nij 03 & EM 02 & Set IV & Set $\eta$ \\ \hline
$c_1$ & $-0.81$ & $-0.76$ & $-0.81$ & $-0.81$ & $-0.81$ \\ \hline
$c_2$ & $ 8.43$ & $ 3.20$ & $ 3.28$ & $ 3.28$ & $ 3.28$ \\ \hline
$c_3$ & $-4.70$ & $-4.78$ & $-3.40$ & $-3.20$ & $-3.80$ \\ \hline
$c_4$ & $ 3.40$ & $ 3.96$ & $ 3.40$ & $ 5.40$ & $ 4.50$ \\ \hline
\end{tabular}
\hspace{4.0mm}
\begin{tabular} {|c|r|}
\hline
LEC & FMS 98 \\ \hline
$\bar d_1+\bar d_2$       & $ 3.06$ \\ \hline
$\bar d_3$                & $-3.27$ \\ \hline
$\bar d_5$                & $ 0.45$ \\ \hline
$\bar d_{14}-\bar d_{15}$ & $-5.65$ \\ \hline
\end{tabular}
\caption{Values for the $\pi$N LECs, from several authors: BM 00 \cite{BM00}, 
Nij 03 \cite{Nij03}, EM 02 \cite{EM02}, Set IV \cite{EM03}, and Set $\eta$ 
\cite{HVA08}. 
The numerical values of the couplings $d_i$'s, which have a small influence 
on NN observables, were taken from FMS 98 \cite{FMS98}.\label{tab1}}
\end{center}
\end{table}

\section{Phase shifts and deuteron properties}

\begin{figure}[!ht]
\begin{center}
\begin{tabular}{ll}
\includegraphics[height=5.3cm]{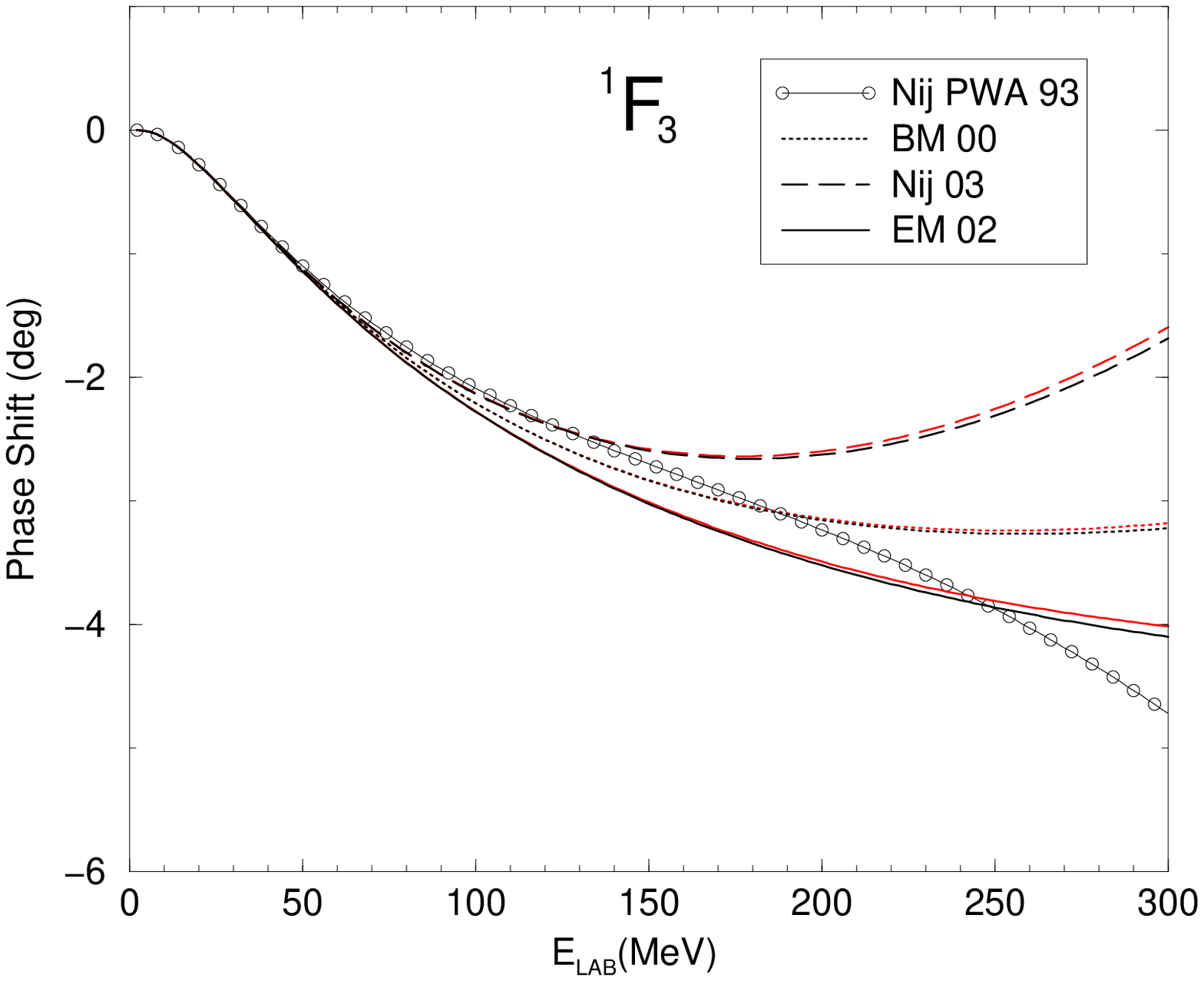}\hspace{5mm}&
\includegraphics[height=5.3cm]{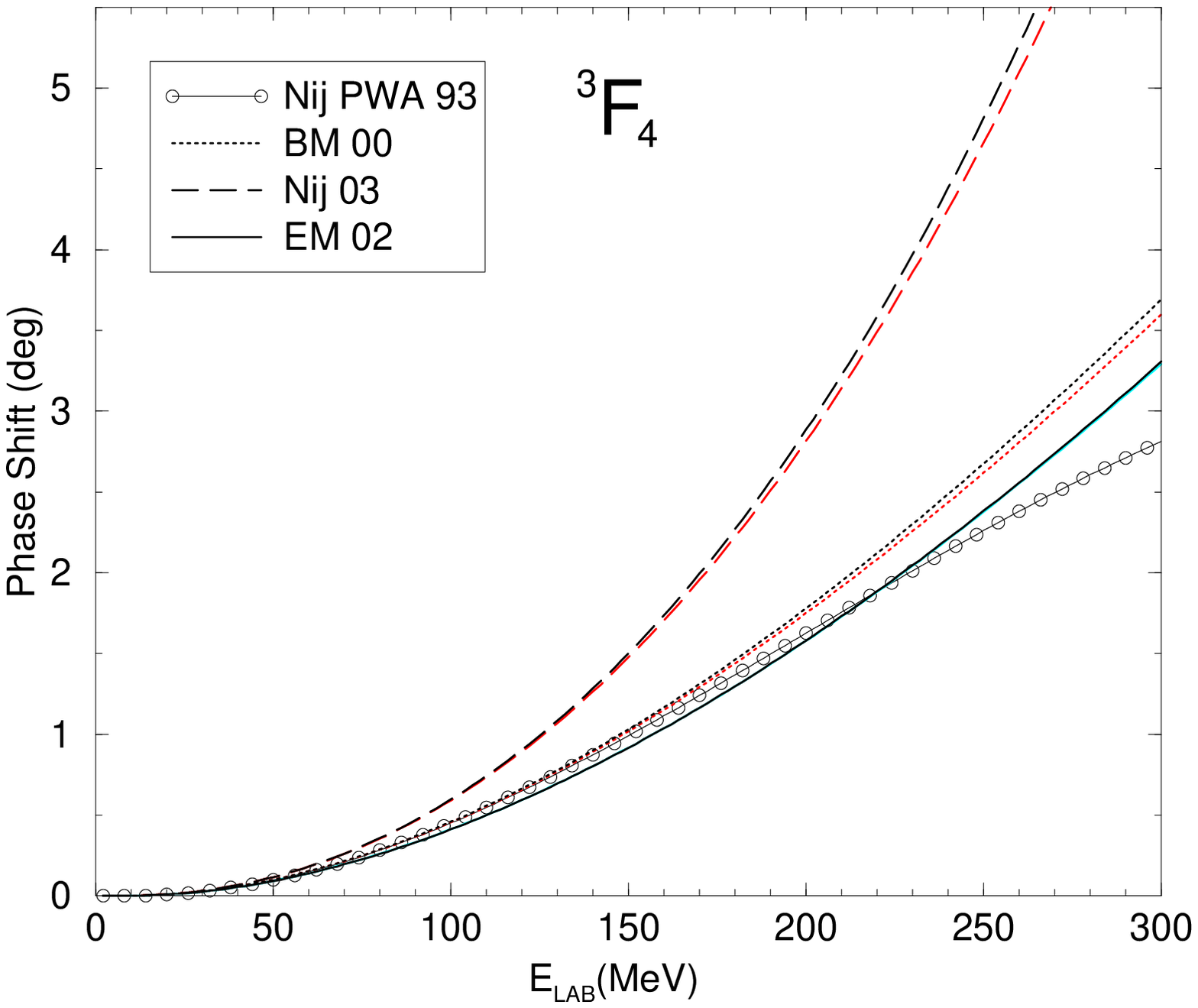}
\end{tabular}
\end{center}
\caption{${}^{1}F_3$ and ${}^{3}F_4$ phase shifts, using LECs from 
Table 1. For comparison, we also plot results from the Nijmegen partial 
wave analysis \cite{nij93} (open circles).}
\label{fig4}
\end{figure}

In Ref. \cite{RH04} phase shifts for some peripheral waves ($F$, $G$, and 
$H$) were computed using the available $\pi$N LECs. The short-distance 
singularities were regulated with the same configuration-space cutoff 
used by the Argonne potentials \cite{argonne}, $[1-\exp(-cr^2)]^4\times$TPEP, 
with $c=2.0{\rm fm}^{-2}$. That work has shown that the differences between 
the RB and HB results for these waves are negligible. That happens due to 
the large superposition of the one-pion exchange potential (OPEP), 
making the difference in Fig.\ref{fig3} hard to observe. 
More pronounced is the dependence of the phase shifts with the set of LECs. 
Fig.\ref{fig4} illustrates that with two selected partial waves. 

Results for lower waves and deuteron properties were calculated in 
Ref. \cite{HVA08} using the renormalization method developed by the 
Granada group \cite{granada}. The central idea of the method is that 
the behavior of the potential at short distances fixes uniquely the 
form of the wave functions at the 
origin\footnote{The short-distance singularity of a potential 
is closely related to the divergent ultraviolet behavior of an 
interaction that is valid only at low energies. In order to have some 
predictive power, such effective theory has to be properly renormalized.}. 
A repulsive potential prevents the two particles to get close to each 
other, which causes the wave function to decrease exponentially in 
magnitude as $r$ goes to zero. In the attractive case, on the other hand, 
physical conditions have to be imposed to avoid the particles to collapse 
at the origin with infinite velocity. In the Granada method, this is 
achieved by demanding the wave function to satisfy specific boundary 
conditions at the origin. There the wave function is specified up to a 
phase, which is determined with an input of a physical quantity. For the 
generalized case of $N$ coupled channels, a potential diverging at the 
origin as 
$
{\bf U} (r) \to {m_{N}{\bf C}_n/r^n} \, , 
$
with ${\bf C}_n$ a matrix of generalized Van der Waals coefficients, 
is diagonalized via an unitary matrix~${\bf G}$
\begin{eqnarray}
m_{N}{\bf C}_n={\bf G}\,{\rm diag}(\pm R_1^{n-2},\dots,\pm R_N^{n-2})\,
{\bf G}^{-1}\,, 
\end{eqnarray} 
with $ R_i$ constants with length dimension. The plus (minus) sign 
corresponds to the case with a positive, attractive (negative, repulsive) 
eigenvalue. At short distances the solutions are given by 
${\bf u}(r)\to{\bf G}\tilde{\bf u}$, where $\tilde{\bf u}$ is a column 
vector with components $u_{1,\pm}(r),\cdots, u_{N,\pm}(r)$.
The attractive and repulsive cases behave respectively as 
\begin{eqnarray}
u_{i,-}(r)&\to&C_{i,-}\left(\frac{r}{R_i}\right)^{n/4}\sin
\left[\frac{2}{n-2}\left(\frac{R_i}{r}\right)^{\frac{n}2-1}+\varphi_i\right]\,,
\label{eq:uA}\\
u_{i,+}(r)&\to&C_{i,+}\left(\frac{r}{R_i}\right)^{n/4}\exp
\left[-\frac{2}{n-2}\left(\frac{R_i}{r}\right)^{\frac{n}2-1}\right]\,.
\label{eq:uR}
\end{eqnarray} 
Here, $\varphi_i$'s are arbitrary short distance phases which 
in general depend on the energy. There are as many short distance 
phases as short distance attractive eigenpotentials. Orthogonality of 
the wave functions at the origin yields the relation 
\begin{eqnarray}
\sum_{i=1}^N\left[{u_{k,i}}^* u_{p,i}'-{u_{k,i}'}^* u_{p,i} \right]\Big|_{r=0}
=\sum_{i=1}^A\cos(\varphi_i(k)-\varphi_i(p))\,,
\end{eqnarray} 
where $A \le N$ is the number of the short distance attractive 
eigenpotentials. 
The $\varphi_i$'s are fixed by low-energy (long-distance) input, which are 
the phase shifts close to threshold. The latter has a well-known 
low-energy expansion, the leading term being the inverse of the scattering 
``lengths''. 
For coupled channels with total (orbital) angular momentum $j$ ($j\pm 1$) 
the threshold behavior in the Stapp-Ypsilantis-Metropolis convention is 
\begin{equation} 
\bar\delta^{1j}_{j-1} \to -k^{2j-1}/\bar a^{1j}_{j-1}\,,\qquad
\bar\delta^{1j}_{j+1} \to -k^{2j+3}/\bar a^{1j}_{j+1}\,,\qquad
\bar\epsilon_j \to  -k^{2j+1}/\bar a^{1j}_{j}\,.
\label{eq:phase-thres}
\end{equation} 

Results for deuteron properties are given in Table \ref{tab2}. For the 
RB potential using Set IV, the asymptotic $D/S$ ratio $\eta$ is a 
prediction, which overshoots the $D$-state contribution. We noted that 
this can be remedied by slightly changing the chiral couplings $c_3$ 
and $c_4$ to reproduce the experimental value of $\eta$ (Set $\eta$). 

\begin{table}[htb]
\begin{center}
\begin{tabular}{|c|c|c|c|c|c|c|c|}
\hline  Set  & $\gamma ({\rm fm}^{-1})$ & $\eta$ & $A_S ( {\rm fm}^{-1/2}) $
& $r_d ({\rm fm})$ & $Q_d ( {\rm fm}^2) $ & $P_D $
\\ \hline\hline
{\rm OPE}  & Input & 0.02634 & 0.8681(1) & 1.9351(5) & 0.2762(1) & 7.88(1)\%
\\ \hline
HB Set IV  & Input & Input & 0.884(4) & 1.967(6) & 0.276(3) & 8(1)\%
\\ \hline
RBE Set IV & Input & 0.03198(3) & 0.8226(5) & 1.8526(10) & 0.3087(2)
& 22.99(13) \%
\\ \hline
RBE Set $\eta$ & Input & 0.02566(1) & 0.88426(2) & 1.96776(1) & 0.2749(1)
& 5.59(1) \%
\\ \hline\hline
NijmII & 0.231605 & 0.02521 & 0.8845(8) & 1.9675 & 0.2707 & 5.635\%
\\
Reid93 & 0.231605 & 0.02514 & 0.8845(8) & 1.9686 & 0.2703 & 5.699\%
\\ \hline \hline
Exp. &  0.231605 &  0.0256(4)  & 0.8846(9)  & 1.971(6)  &
0.2859(3) & ---
\\ \hline
\end{tabular}
\caption{Deuteron properties, for several NN potentials. \cite{HVA08}.
\label{tab2}}
\end{center}
\end{table}

\begin{figure}[!ht]
\begin{center}
\begin{tabular}{ll}
\includegraphics[height=4.5cm,width=6.0cm]{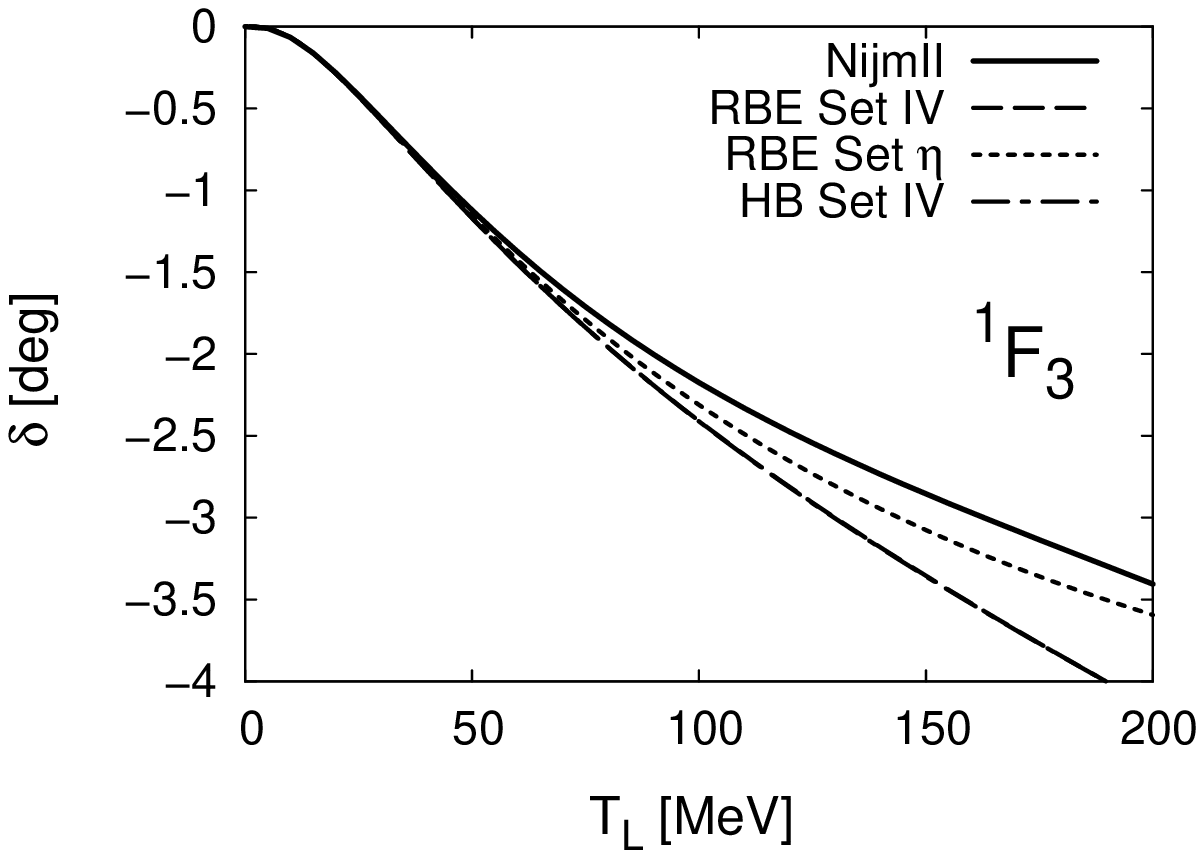}\hspace{1mm}&
\includegraphics[height=4.5cm,width=6.0cm]{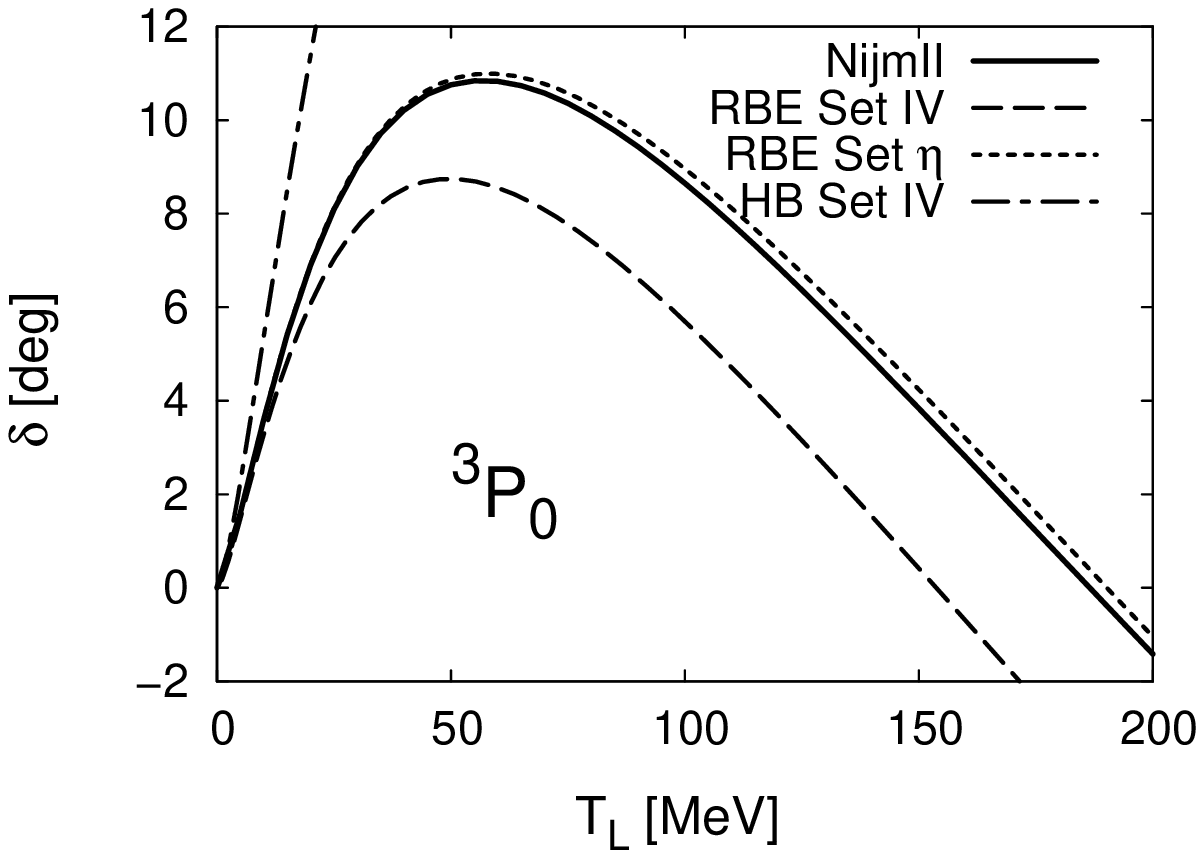}\\
\includegraphics[height=4.5cm,width=6.0cm]{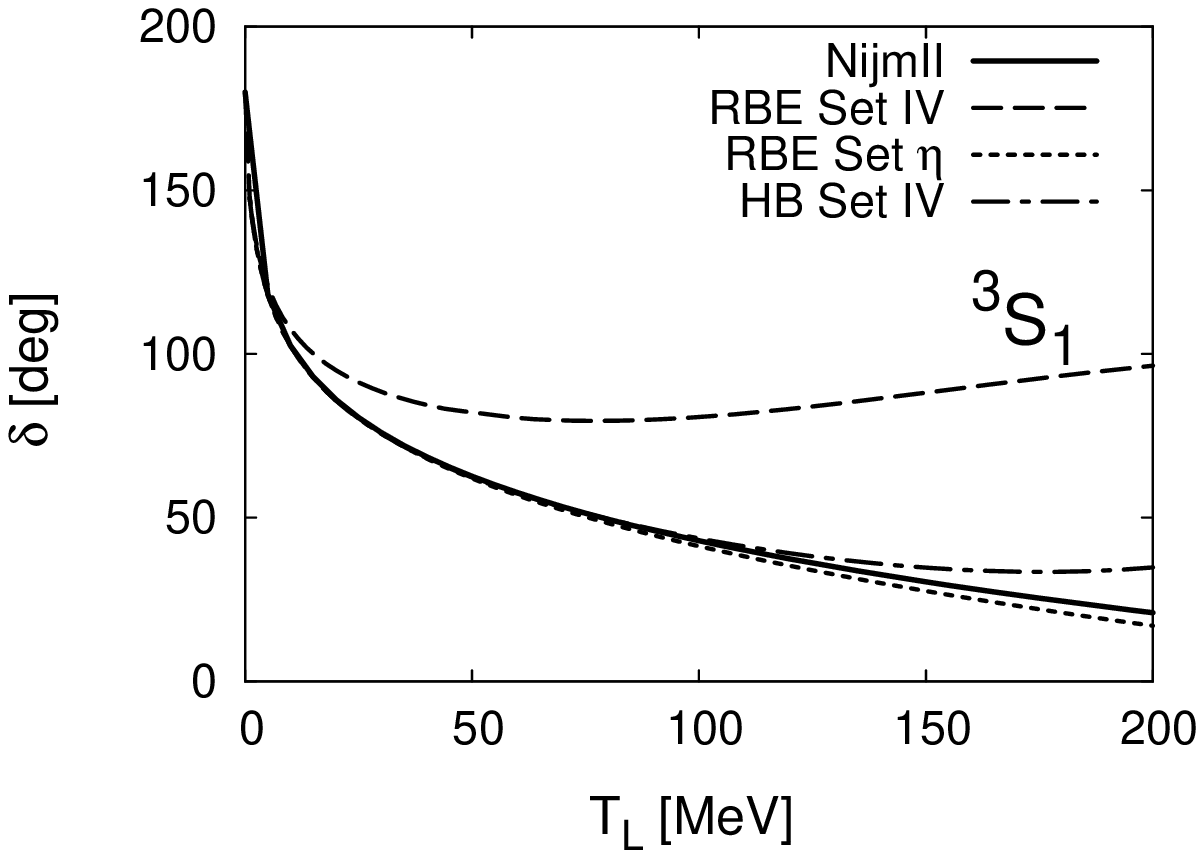}\hspace{1mm}&
\includegraphics[height=4.5cm,width=6.0cm]{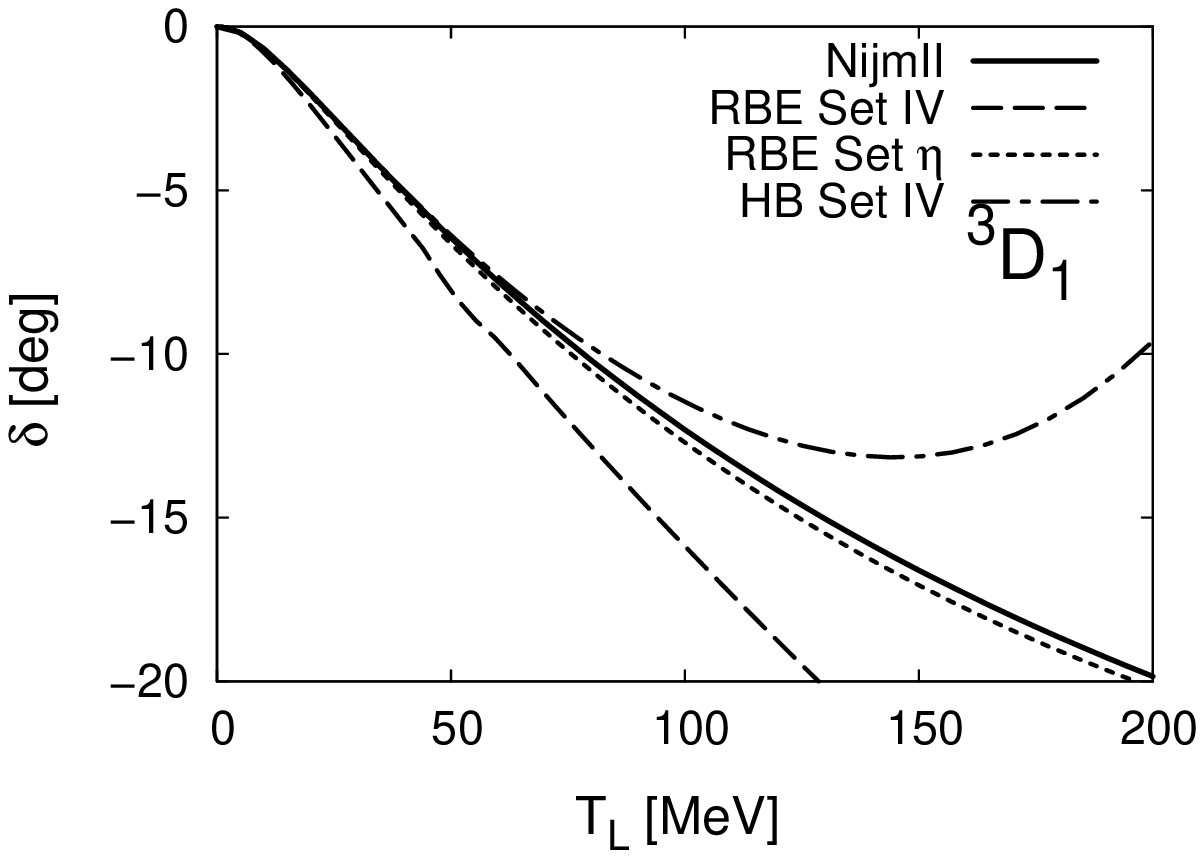}
\end{tabular}
\end{center}
\caption{Selected partial wave phase shifts using Sets IV and $\eta$, 
compared with the HB results (Set IV) and the Nijmegen partial wave 
analysis \cite{nij93}.}
\label{fig5}
\end{figure}

A surprising result of the RB potential with Set $\eta$ is that the number 
of required low-energy inputs to renormalize the Schr\"odinger equation is 
significantly less than in the HB case, nearly a half (see 
Table II of Ref. \cite{HVA08}). This is a consequence of the different 
behavior of the RB potential near the origin, $V(r)\sim 1/r^7$, which 
resembles a relativistic Wan der Waals force, in contrast with the 
typical $1/r^6$ of the HB counterpart. 
There is also see a clear improvement in the deuteron properties and 
on the energy dependence of several phase shifts, some of them shown in 
Fig. \ref{fig5}. There are, however, some partial waves (${}^{1}S_{0}$, 
${}^{1}D_{2}$, ${}^{3}F_{3}$, and ${}^{3}P_{2}$) where 
agreement with the Nijmegen partial wave analysis gets slightly worse. 
Compared to the HB, the ${}^{1}S_{0}$, ${}^{3}P_{2}$, $\epsilon_{2}$, 
${}^{3}D_{3}$, and ${}^{3}G_{5}$ RB results are not satisfactory, but 
in other cases they are either consistent or better, remarkably for 
${}^{1}D_{2}$, ${}^{3}P_{0}$, ${}^{3}P_{1}$, ${}^{3}D_{1}$, and 
$\epsilon_{1}$ \cite{HVA08}. 

\section{Summary}

The program of constructing a nucleon-nucleon interaction based on the 
ideas of Becher and Leutwyler was presented. We pointed out the 
problem with the usual HB formalism in the one-nucleon sector 
and its consequences to the nucleon-nucleon system, namely, that it 
cannot account for the correct asymptotic description of the TPEP. 
If one performs a naive $1/m_N$ expansion of our results (which is not 
correct) one should, in principle, recover the HB expressions. In doing 
so, we noticed that there are three discrepant terms coming from 
two-loop dynamics. The origin of this discrepancy is not yet known and 
should be resolved if more precision calculations are demanded. 

Considering only the effects of the $1/m_N$ expansion in the covariant 
loop functions, we showed that the HB and RB versions of the TPEP can 
have considerable differences, in particular the partial waves with 
total isospin $T=0$ in the physically interesting region between 2 and 5fm. 
However, this effect is barely noticeable in peripheral phase shifts, 
due to the large one-pion exchange contribution. More pronounced is their 
sensitivity to the $\pi$N LECs, in particular, $c_3$ and $c_4$. 

To obtain the RB predictions for lower partial waves and deuteron properties, 
we used the renormalization method developed by the Granada group. We 
readjusted the LECs $c_3$ and $c_4$ to reproduce the asymptotic $D/S$ ratio 
(Set $\eta$), bringing other deuteron properties in agreement with 
experimental data and other potential model predictions. We also obtained 
an overall good agreement of the calculated phase shifts with the Nijmegen 
partial wave analysis, in a similar way as in the HB case. However, 
such agreement was obtained with significantly less counterterms (nearly a 
half) than the HB potential. That could be an advantage in determining 
the nucleon-nucleon LECs and bringing the RB potential into a form that 
could be eventually used in nuclear calculations.

\acknowledgments 

I would like to thank Manoel R. Robilotta, Carlos A. da Rocha, Manolo 
P. Valderrama and Enrique R. Arriola for collaboration and interest on 
this project, and to the conveners and organizers for the enjoyable and 
well-organized conference, and the opportunity to present this talk.


\end{document}